\documentclass[12pt]{iopart}

\usepackage{iopams}
\usepackage{braket}
\usepackage{amsfonts}
\usepackage{color}
\usepackage[dvipsnames]{xcolor}
\usepackage{amssymb}
\usepackage{mathrsfs}
\usepackage{graphicx}
\usepackage{lmodern}
\usepackage{lineno}
\usepackage{hyperref}
\usepackage[normalem]{ulem}

\newcommand{\beq}{\begin{eqnarray}}
\newcommand{\eeq}{\end{eqnarray}}

\newcommand{\Cyl}{\mathrm{Cyl}}

\def\keywords#1{\vspace{10pt}
     \begin{indented}
     \item[]\rm Keywords: #1\par
     \end{indented}}

\begin{document}



\title{Quasi-probability distributions in Loop Quantum Cosmology}
\author{Jasel Berra--Montiel$^{1,2}$ and 
Alberto Molgado$^{1,2}$}

\address{$^{1}$ Facultad de Ciencias, Universidad Aut\'onoma de San Luis 
Potos\'{\i} \\
Campus Pedregal, Av. Parque Chapultepec 1610, Col. Privadas del Pedregal, San
Luis Potos\'{\i}, SLP, 78217, Mexico}
\address{$^2$ Dual CP Institute of High Energy Physics, Mexico}

\eads{\mailto{\textcolor{blue}{jasel.berra@uaslp.mx}},\ 
\mailto{\textcolor{blue}{alberto.molgado@uaslp.mx}}\ 
}


\begin{abstract}

In this paper, we introduce a complete family of parametrized quasi-probability distributions in phase space and their corresponding  Weyl quantization maps with the aim to generalize the recently developed Wigner-Weyl formalism within the Loop Quantum Cosmology program (LQC).  In particular, we intend to define those quasi-distributions for states valued on the Bohr compactification of the real line in such a way that they are labeled by a parameter that accounts for the ordering ambiguity corresponding to non-commutative quantum operators.   Hence, we notice that the projections of the parametrized quasi-probability distributions result in marginal probability densities which are invariant under any ordering prescription.
We also note that, in opposition to the standard Schr\"odinger representation, for an arbitrary character the quasi-distributions determine a positive function independently of the ordering.  Further, by judiciously implementing a parametric-ordered Weyl quantization map for LQG, we are able to recover in a simple manner the relevant cases of the standard, anti-standard, and Weyl symmetric orderings, respectively.  
We expect that our results may serve to analyze several fundamental aspects within the LQC program, in special those related to coherence, squeezed states, and the convergence of operators, as extensively analyzed in the quantum optics and in the quantum information frameworks.

\end{abstract}

\keywords{Loop Quantum Cosmology, quasi-probability distributions}
\ams{83C45, 81S30, 46F10}


\section{Introduction}

In General Relativity, the gravitational field is intrinsically described by the geometry of the space-time. As a consequence, most of the distinguishing features of gravity, for instance, the formation of gravitational waves, black holes, the expansion of the Universe, among others, may be directly related to geometrical properties. However, although the successful portrait of the dynamical nature of space-time provided by  Einstein theory, the emergence of classical singularities prevents predictability even on physically well-posed situations \cite{Hawking}. In order to overcome this kind of difficulties, a mathematically consistent quantum theory for the gravitational field results imperative. This issue is addressed within the so-called Loop Quantum Gravity program (LQG),
as described in \cite{Rovelli}, \cite{Ashtekar}, \cite{Thiemann}, \cite{Rovelli2}. One of the essential elements of LQG is the use of non-perturbative methods, as those developed for non-Abelian gauge field theories, to the gravitational field in a fully covariant manner. The introduction of this quantization formalism for symmetric cosmological models such as isotropic and homogeneous space-times led to the Loop Quantum Cosmology  
framework (LQC)~\cite{BojowaldLQC}. In this direction, significant advances in the quantum gravity program have been accomplished within the cosmological context, remarkably including the avoidance of classical singularities through a quantum bounce \cite{BojowaldB}, a macroscopic ground for the black hole entropy \cite{RovelliBH}, \cite{Baez}, \cite{Domagala}, and inhomogeneities imprinted on the cosmic expansion \cite{Agullo}. Despite these critical results, a proper understanding of the semiclassical limit of the quantum regime remains an open subject. For cosmological scenarios, the usual approach to handle this situation is based on analyzing minisuperspace models, that is, phase space reduced solutions where all the kinematical symmetries have been fixed, for instance, by means of homogeneity conditions, thereby making some progress in the LQC program~\cite{Singh}. In addition, it is important to mention that LQC, also referred in the literature as the polymer quantization in the non-cosmological context, has been widely used to explore both mathematical and physical properties of background-independent theories \cite{Tecotl1}, \cite{Tecotl2}, \cite{Velhinho}, \cite{Corichi}, \cite{Riemann}.

With the aim to shed some light on the issues mentioned above, and following some ideas introduced by Cahill and Glauber in \cite{Cahill}, in this manuscript we intend to describe the LQC program by means of quasi-probability distributions defined in  phase space. The first quasi-probability distribution was introduced almost 90 years ago by Wigner in order to calculate quantum corrections to the classical Boltzmann distribution associated to quantum mechanical systems \cite{Wigner}, and since then, a complete family of different quasi-probability distributions has been analyzed from several physical perspectives, including quantum optics and quantum information \cite{Scully}, \cite{Nielsen}, \cite{Weinbub}, establishing phase space quantization as a valuable technique to study a huge variety of dynamical and semiclassical aspects of quantum theory (see~\cite{Zachos}, and references therein). Within the LQC and LQG scenarios, the only quasi-probability distribution that has been recently studied is the Wigner function, specifically in reference \cite{Sahlmann}, where the Wigner function over the Bohr compactification of the real line was defined.  Further, its relation to the Schr\"odinger representation as a limit for systems with finite and infinite degrees of freedom was determined in \cite{PolyDQ}, \cite{PolyW}. Being our main aim to extend these results, the purpose of the paper is to obtain a complete family of $s$-parametrized quasi-probability distributions for LQC and their corresponding $s$-ordered Weyl quantization map. Determining these quasi-probability distributions may allow us to analyze in a more concise manner several aspects related to correlations functions, squeezed states, and the correspondence between the classical and quantum coherence theory, which are fundamental issues to understand within the LQC framework.  We also hope that the introduction of these distributions will be beneficial in LQC by suitably interpreting well-known features previously developed in either the quantum optics or the quantum information theory frameworks.

The paper is organized as follows: In section 2 we briefly introduce the basic concepts behind the Weyl-Wigner quantization scheme for any ordering prescription. In section 3, the $s$-parametrized quasi-probability distributions for LQC are determined. Then, by employing some tools of harmonic analysis associated with the Bohr compactification of the reals, we introduce the $s$-ordered Weyl quantization map for LQC. Finally, we present some concluding remarks in section 4.

\section{The $s$-ordered Wigner-Weyl quantization}
In this section, we briefly review the general Wigner-Weyl quantization procedure for different orderings of non-commutative operators and their corresponding family of quasi-probability distributions in phase space. The formulation presented is closely related to the deformation quantization picture of quantum mechanics based on the introduction of an associative and non-commutative product known as the star product on the algebra of classical observables. For a more complete exposition regarding the general $s$-ordering prescription in phase space quantum mechanics we refer the reader to the original papers \cite{Cahill}, \cite{Agarwal}, and also references \cite{Manko}, \cite{Soloviev} for the star product approach. For the sake of simplicity, we restrict our analysis to systems with one degree of freedom, however, a generalization to systems with more dimensions may be built in a straightforward manner.

\subsection{The $s$-ordered Wigner-Weyl correspondence and $s$-parametrized quasi-probability distributions}

One of the principal axioms of quantum mechanics is related to the construction of a map between arbitrary real-valued functions $f$ defined on the classical phase space $\mathbb{R}^{2}$ and self-adjoint operators $\hat{f}$ on the quantum Hilbert space    $L^{2}(\mathbb{R})$,
in such a manner that the algebraic properties of the classical functions are, in some measure, preserved by the quantum operators. The general quantization procedure, known as  Weyl quantization \cite{Weyl}, \cite{Moyal}, satisfies most of the required properties, even though, as the Groenewold's theorem states \cite{Groenewold}, the issue of the ordering ambiguity has to be solved.  With the aim to unify all the possible ordering outcomes, Cahill and Glauber introduced the so-called parametric ordering prescription \cite{Cahill}, where a parameter $s$ varies continuously in the interval $[-1,1]$, thus changing from normal  to antinormal ordering while remaining Weyl symmetric for the value $s=0$. This parametrization has been proven to be a reliable tool in order to analyze the convergence of  operators in terms of holomorphic variables, within the generalized star product representation \cite{Berezin}, \cite{Soloviev1}, and particularly it allows us to introduce different quasi-probability distributions over the phase space, which are extensively used in quantum optics and quantum information \cite{Scully}, \cite{Perelomov}, \cite{Ibort}, \cite{CSDQ}.

Let $f\in \mathcal{S}'(\mathbb{R}^{2})$ be an arbitrary classical function on the phase space $\mathbb{R}^{2}$, where $\mathcal{S}'(\mathbb{R}^{2})$ denotes the Schwartz space of tempered distributions, that is, the space of continuous linear functionals on  the Schwartz space of rapidly decreasing smooth functions, $\mathcal{S}(\mathbb{R}^{2})$, \cite{Reed}. We define the $s$-ordered Weyl quantization of $f$, as an operator on the Hilbert space $L^{2}(\mathbb{R})$, by (in natural units where $\hbar=1$)
\begin{equation}\label{sWeyl}
\mathcal{Q}_{s}(f)=\frac{1}{4\pi^{2}}\int_{\mathbb{R}^{2}}d\lambda\,d\mu\,\tilde{f}(\lambda,\mu)e^{-is\lambda\mu/2}e^{i(\lambda \hat{q}+\mu\hat{p})}, 
\end{equation}
where $\tilde{f}(\lambda,\mu)$ stands for the Fourier transform given by
\begin{equation}\label{Fourier}
\tilde{f}(\lambda,\mu)=\int_{\mathbb{R}^{2}}dq\,dp\,f(q,p)e^{-i(\lambda q+\mu p)},
\end{equation}
$s\in [-1,1]$ denotes a continuous parameter, and the position and momentum operators $\hat{q}, \hat{p}\in \mathcal{L}(L^{2}(\mathbb{R}))$ satisfy the usual canonical commutation relations
\begin{eqnarray}\label{ccr1}
[\hat{q},\hat{p}]\psi &=& i\psi,\\
\left[ \hat{q},\hat{q}\right] \psi &=&0=\left[\hat{p},\hat{p} \right]\psi, \label{ccr2} 
\end{eqnarray}
for any vector $\psi\in L^{2}(\mathbb{R})$. We may  regard expression (\ref{sWeyl}) as a Bochner integral for functions with values in a Banach space but, for our analysis, it turns out to be more convenient to consider operators on the Hilbert space $L^{2}(\mathbb{R})$ as integral operators using similar properties of those of the Lebesgue integral \cite{Yosida}. By substituting the Fourier transform (\ref{Fourier}) into (\ref{sWeyl}) one obtains
\begin{equation}\label{squantization}
\hat{f}=\mathcal{Q}_{s}(f)=\frac{1}{2\pi}\int_{\mathbb{R}^{2}}dq\,dp\,f(q,p)\hat{\Delta}_{s}(q,p),
\end{equation}
where the operator $\hat{\Delta}_{s}(q,p)$ reads
\begin{equation}
\hat{\Delta}_{s}(q,p)=\frac{1}{2\pi}\int_{\mathbb{R}^{2}}d\lambda\,d\mu\,e^{-is\lambda\mu/2}e^{-i\lambda(q-\hat{q})-i\mu(p-\hat{p})}.
\end{equation}
The operator $\hat{\Delta}_{s}(q,p)$ corresponds to the $s$-ordered version of the Weyl-Stratonovich quantizer \cite{Stratonovich}, and satisfies the following relations
\begin{eqnarray}
\hat{\Delta}^{\dagger}_{s}(q,p)=\hat{\Delta}_{-s}(q,p), \\
\tr\left\lbrace\hat{\Delta}_{s}(q,p) \right\rbrace=1, \\ 
\tr\left\lbrace \hat{\Delta}_{s}(q,p)\hat{\Delta}^{\dagger}_{s}(q',p') \right\rbrace=2\pi\delta(q-q')\,\delta(p-p'). 
\end{eqnarray}  
By employing the properties of $\hat{\Delta}_{s}(q,p)$ and the Baker-Campbell-Hausdorff formula, it is possible to express the $s$-ordered Weyl quantization in its integral 
representation given by 
\begin{equation}\label{qintegral}
\mkern-20mu\mathcal{Q}_{s}(f)\psi(q)=\frac{1}{2\pi}\int_{\mathbb{R}^{2}}dy\,dp\,f\left(\frac{1}{2}y(1-s)+\frac{1}{2}q(1+s),p\right)e^{-i(y-q)p}\psi(y). 
\end{equation}
This integral may not necessarily be  absolutely convergent and it must be understood as the result of computing a partial Fourier transform with respect to the second variable. By Plancherel formula and Fubini's theorem, the partial Fourier transform occurs to be an isometry on $L^{2}(\mathbb{R}^{2})$ and, in consequence, it extends to a unitary map of $L^{2}(\mathbb{R}^{2})$ to itself, and thus the partial Fourier transform may be computed by employing the usual formulas on functions defined in $L^{1}(\mathbb{R}^{2n})\cap L^{2}(\mathbb{R}^{2n})$ \cite{Reed}.
As an example, let us select as the classical observable the monomial $q^m p^n\ (m,n\in\mathbb{Z}^+)$, by using the $s$-ordered Weyl quantization prescription as defined in (\ref{sWeyl}), this function is mapped into the operator $\hat{q}^m \hat{p}^n$, when $s=1$, and into $\hat{p}^n q^m$ when $s=-1$, i.e., the ordering specified by the values $s=1$ and $s=-1$ correspond to the standard and the anti-standard orderings, respectively. In addition, the Weyl-symmetric ordering is obtained by taking the value $s=0$ \cite{Soloviev}. 

Now, given an operator $\hat{f}\in\mathcal{L}(L^{2}(\mathbb{R}))$, let us define 
its corresponding  $s$-ordered phase space function, also known as its symbol, according to the terminology adopted in harmonic analysis \cite{Folland}, by 
\begin{equation}\label{sdequantize}
W_{\hat{f}}^{s}(q,p)=\tr\left\lbrace \hat{f}\hat{\Delta}_{s}(q,p)\right\rbrace.
\end{equation}
This formula allows us to obtain the inverse relation of the $s$-ordered Weyl quantization map (\ref{sWeyl}), that is, for each operator $\hat{f}$ it is possible to determine a function $W_{\hat{f}}^{s}(q,p)$ on the phase space $\mathbb{R}^{2}$, such that the composition satisfies $\mathcal{Q}_{s}(W_{\hat{f}}^{s})=\hat{f}$;  for this reason the relation (\ref{sdequantize}) is also known as the $s$-ordered dequantizer in the literature \cite{Marmo}.

To finish this section, let us compute the $s$-ordered symbol associated with the density operator $\hat{\rho}=\ket{\psi}\bra{\psi}$ of a given quantum state $\psi\in L^{2}(\mathbb{R})$. By means of the formula (\ref{sdequantize}), we obtain
\begin{eqnarray}\label{sWigner}
\rho_{s}(\psi)(q,p)&&:=W^{s}_{\hat{\rho}}(q,p)=\tr\left\lbrace \hat{\rho}\hat{\Delta}_{s}(q,p)\right\rbrace, \nonumber\\
&&=\int_{\mathbb{R}}dy\,\overline{\psi}\left( q-\frac{y}{2}(1-s)\right) \psi\left( q+\frac{y}{2}(1+s)\right) e^{-iyp}.
\end{eqnarray}
By employing the definition of the Fourier transform (\ref{Fourier}), this symbol may be also expressed in terms of the momentum basis as
\begin{equation}\label{sWignerm}
\rho_{s}(q,p)=\frac{1}{2\pi}\int_{\mathbb{R}}dk\,\overline{\widetilde{\psi}}\left( p-\frac{k}{2}(1+s)\right) \tilde{\psi}\left(p+\frac{k}{2}(1-s) \right)e^{ikq} \,, 
\end{equation}
where $\tilde{\psi}$ stands for a quantum state in the momentum space.

It is important to mention here that a different representation for the symbol of the density operator arises  by using the holomorphic representation instead of the position and momentum variables. Indeed, if we substitute $q=(\alpha+\overline{\alpha})/\sqrt{2}$ and $p=(\alpha+\overline{\alpha})/i\sqrt{2}$ in equation (\ref{sWigner}), we obtain the most familiar expression of the $s$-parametrized quasi-probability distribution
\begin{equation}
\rho_{s}(\psi)(\alpha)=\frac{1}{\pi^{2}}\int_{\mathbb{R}^{2}}d^{2}\beta\,\tr\left\lbrace \hat{\rho}\hat{D}(\beta)\right\rbrace e^{\alpha\overline{\beta}-\overline{\alpha}\beta+s|\beta|^{2}/2}, 
\end{equation}  
where $\hat{D}(\beta)$ corresponds to the Glauber displacement operator given by $\hat{
D}(\beta)=e^{\beta\hat{a}^{\dagger}-\overline{\beta}\hat{a}}$ while the operators $\hat{a}^{\dagger}$ and $\hat{a}$ stand for the standard creation and annihilation operators, respectively \cite{Gluaber}. 
For the case $s=1$, the expression for $\rho_{s}(\psi)(\alpha)$ reduces to the Glauber-Sudarshan $P$-function (normal ordering), for the value $s=0$, we obtain the Wigner function (Weyl-symmetric ordering), and for $s=-1$ we have the Husimi $Q$-distribution (antinormal ordering) \cite{Scully}. These particular cases of the generalized $s$-ordered quasi-probability distributions are extensively used in quantum optics and quantum information, as they provide enough information in order to completely characterize a quantum system by displaying the interference effects associated to quantum states, and by aiding to identify negative values on some regions of the phase space for strongly non-classical behaviors. Furthermore, it happens that it is possible to reconstruct various quasi-probability distributions direct from experimental data, a method known as quantum-state tomography \cite{Ibort}, \cite{Vogel}, \cite{TDQ}. 

\section{Quasi-probability distributions in LQC}

\subsection{$s$-parametrized quasi-probability distributions in LQC}

In this section, we define the $s$-parametrized quasi-probability distributions for states with values on $b\mathbb{R}$, where $b\mathbb{R}$ stands for the Bohr compactification of the reals; also the $s$-ordered quantization map for LQC is obtained. First, we start with a brief review of some basic features concerning the analytical properties of functions defined on $b\mathbb{R}$ (we refer the reader to \cite{Harmonic}, \cite{Reiter} for further details).

Let $\mathbb{R}$ be the group of real numbers, equipped with the usual addition operation and the standard topology. A character $h_{\mu}$ of $\mathbb{R}$ is a group homomorphism $h_{\mu}:\mathbb{R}\to\mathbb{T}$, to the unit torus, with $\mathbb{T}=\left\lbrace z\in\mathbb{C}:|z|=1\right\rbrace $, so $h_{\mu}$ is a map satisfying
\begin{equation}
h_{\mu}(b+c)=h_{\mu}(b)h_{\mu}(c),
\end{equation} 
for every $b,c\in \mathbb{R}$. The set of all characters $h_{\mu}$ of $\mathbb{R}$, labeled by $\mu\in\mathbb{R}$,
\begin{equation}
h_{\mu}(c)=e^{i\mu c},
\end{equation}
forms a group, called the dual group of $\mathbb{R}$, which proves to be isomorphic to the reals and will be denoted by $\widehat{\mathbb{R}}$. Let us consider now  the dual group of the reals but equipped with the discrete topology, $\widehat{\mathbb{R}}_{\mathrm{discr}}$. By definition, this group is discrete and Abelian, then by the Pontryagin's duality theorem, its dual group $\widehat{\widehat{\mathbb{R}}}_{\mathrm{discr}}$ forms a compact Abelian group, the so-called Bohr compactification of the reals $b\mathbb{R}$. Furthermore, the group $\mathbb{R}$ can be embedded as a dense subgroup of $b\mathbb{R}$ and the addition operation in $\mathbb{R}$ extends uniquely to the continuous group operation of $b\mathbb{R}$, given by the characters.

Since, both $b\mathbb{R}$ and $\widehat{b\mathbb{R}}$ are locally compact Abelian groups, they carry a unique and normalized Haar measure given by,
\begin{equation}\label{haardRb}
\int_{\widehat{b\mathbb{R}}}d\mu\,\tilde f_{\mu}=\sum_{\mu\in\mathbb{R}}\tilde{f}_{\mu},
\end{equation} 
that is, the Haar measure $d\mu$ on $\widehat{b\mathbb{R}}$ corresponds to the counting measure on $\mathbb{R}$. On the other hand, the measure on $b\mathbb{R}$ satisfies
\begin{equation}\label{haarRb}
\int_{b\mathbb{R}}dc\,h_{\mu}(c)=\delta_{\mu,0}.
\end{equation}
for any character $h_{\mu}$. Here, $\tilde{f}_{\mu}$ denotes the Fourier transform on $b\mathbb{R}$, i.e., an isomorphism between $L^{2}(b\mathbb{R},da)$  and $L^{2}(\widehat{b\mathbb{R}},d\mu)$ such that
\begin{equation}
\tilde{f}_{\mu}=\int_{b\mathbb{R}}dc\,f(c)h_{-\mu}(c).
\end{equation}
Given the discrete topology endowed in $\widehat{b\mathbb{R}}$, and the Peter-Weyl theorem, the characters $h_{\mu}$ comprise an orthonormal uncountable basis for $L^{2}(b\mathbb{R},da)$, yielding a non-separable Hilbert space.

In order to establish  the family of $s$-parametrized quasi-probability distributions for LQC, and following the terminology developed in \cite{Sahlmann}, let us denote the set of cylindrical functions by $\Cyl(b\mathbb{R})$, i.e. the finite span of characters defined on $b\mathbb{R}$, and by $\Cyl(\widehat{b\mathbb{R}})$ the image of $\Cyl(b\mathbb{R})$ under the Fourier transform. This implies that any $\psi\in\Cyl(b\mathbb{R})$ can be written in the form
\begin{equation}\label{sum}
\psi(c)=\sum_{\mu\in\mathbb{R}}\tilde{\psi}_{\mu}h_{\mu}(c),
\end{equation}
where $\tilde{\psi}_\mu$, stands for the Fourier coefficient
\begin{equation}
\tilde{\psi}_{\mu}=\int_{b\mathbb{R}}dc\,\psi(c)\overline{h_{\mu}(c)}.
\end{equation}
The expression depicted for $\psi$ in terms of a discrete sum (\ref{sum}), is nonzero only for a countable number of points $\mu\in\mathbb{R}$. Consequently, the space $\Cyl(\widehat{b\mathbb{R}})$ is given by all complex valued functions on $\mathbb{R}$ that vanish at all but a countable number of points.

Next, let us emphasize the role involved by the canonical commutation relations (\ref{ccr1}) and (\ref{ccr2}), on the determination of the quantization map. While in the Schr\"odinger representation of quantum mechanics, the position and momentum operators $\hat{q}, \hat{p}$ are well defined operators on the Hilbert space $L^{2}(\mathbb{R})$, in LQC this is no longer true. A fundamental difference between the Schr\"odinger and the LQC representations lies in the fact that on $L^{2}(b\mathbb{R})$ does not exist an Hermitian operator corresponding to the position operator \cite{Bojowald}, \cite{Tecotl}. Instead, one defines the operators $\hat{h}_{\mu}$ and $\hat{p}$ by
\begin{equation}
\hat{h}_{\mu}\psi(c)=h_{\mu}(c)\psi(c), \;\;\; \hat{p}\psi(c)=\sum_{\mu\in\mathbb{R}}\mu\,\tilde{\psi}_{\mu}h_{\mu}(c),
\end{equation}   
where $\psi\in\Cyl(b\mathbb{R})$. Since, $h_{\mu}(c)\psi(c)$ proves to be non differentiable on $L^{2}(b\mathbb{R})$, this precludes the existence of a position operator on the Hilbert space $L^{2}(b\mathbb{R})$.

With the aim to generalize the $s$-ordered quasi-probability distribution obtained in (\ref{sWigner}) to the Bohr compactification of the real line,  we will consider the momentum representation (\ref{sWignerm}), in agreement with reference~\cite{Sahlmann}, since there is no general consensus on how to multiply an element of $b\mathbb{R}$ by an arbitrary real number. Bearing this in mind, we define the $s$-parametrized family of quasi-probability distributions for a state $\psi\in \Cyl(b\mathbb{R})$, as the complex valued function on $b\mathbb{R}\times\widehat{b\mathbb{R}}$ by
\begin{equation}\label{WignerLQC}
\rho^{LQC}_{s}(\psi)(c,\mu):=\int_{\widehat{b\mathbb{R}}}d\nu\,\overline{\tilde{\psi}}_{\mu-(1+s)\nu/2}\tilde{\psi}_{\mu+(1-s)\nu/2}h_{\nu}(c). 
\end{equation}
We observe that in the case when the parameter $s=0$, the function $\rho^{LQC}_{s}(\psi)(c,\mu)$ corresponds to the Wigner distribution for LQC \cite{Sahlmann}, \cite{PolyDQ}. By using the integration formulas (\ref{haardRb}) and (\ref{haarRb}), the projections of the $s$-parametrized quasi-probability distribution on the Bohr compactification and its dual, respectively, result in the well-known marginal probability densities
\begin{eqnarray}
\int_{b\mathbb{R}}dc\,\rho^{LQC}_{s}(c,\mu)=|\tilde{\psi}_{\mu}|^{2}, \\
\int_{\widehat{b\mathbb{R}}}d\mu\rho^{LQC}_{s}(c,\mu)=|\psi(c)|^{2},
\end{eqnarray}
for all $s\in [-1,1]$, thus indicating that this property is invariant under any ordering prescription. Since any wavefunction on the configuration space,   $\psi\in\Cyl(b\mathbb{R})$, can be written in terms of the characters (\ref{sum}), let us compute the $s$-ordered quasi-probability distribution associated to a single character $h_{\mu'}$. By substituting $\psi(c)=h_{\mu'}(c)$ on (\ref{WignerLQC})
\begin{equation}
\rho^{LQC}_{s}(h_{\mu'})(c,\mu)=\overline{\tilde{\psi}}_{\mu'}\tilde{\psi}_{\frac{2}{1+s}\mu-\frac{1-s}{1+s}\mu'}h_{\frac{2}{1+s}(\mu-\mu')},
\end{equation}
we observe that the second factor in this expression is zero except if $\frac{2}{1+s}\mu-\frac{1-s}{1+s}\mu'=\mu'$, that is, if $\mu=\mu'$. Then
\begin{equation}
\rho^{LQC}_{s}(h_{\mu'})(c,\mu)=|\tilde{h}_{\mu'}|^{2}h_{0}(c)\delta_{\mu',\mu}=\delta_{\mu',\mu}\geq 0.
\end{equation}
This means that, given a character $h_{\mu}:b\mathbb{R}\to\mathbb{T}$, the corresponding quasi-distribution $\rho^{LQC}_{s}(h_{\mu})(c,\mu)$  determines a positive function on $\Cyl(b\mathbb{R}\times\widehat{b\mathbb{R}})$ for any ordering. This feature seems to contradict  Hudson's theorem, which states that the only positive quasi-probability distributions are provided by Gaussian functions~\cite{Hudson}. Nevertheless, as mentioned in \cite{Sahlmann}, this peculiarity is related to the fact that on $b\mathbb{R}$ the characters or plane waves are indeed normalizable and become an orthonormal basis for the Hilbert space $L^{2}(b\mathbb{R})$, contrary to the situation in the standard Schr\"odinger representation.

\subsection{The $s$-ordered Wigner-Weyl correspondence in LQC}

In order to derive the $s$-ordered Weyl quantization map for LQC, in complete analogy to the case depicted in (\ref{squantization}), let us consider a distribution $g$ on $\Cyl(b\mathbb{R}\times\widehat{b\mathbb{R}})$, that is, $g\in\Cyl(b\mathbb{R}\times\widehat{b\mathbb{R}})^{*}$, where $\Cyl(b\mathbb{R}\times\widehat{b\mathbb{R}})^{*}$ denotes the dual space of $\Cyl(b\mathbb{R}\times\widehat{b\mathbb{R}})$~\cite{Sahlmann}. As mentioned before, with the aim to avoid the issue of multiplying an element of $b\mathbb{R}$ by any real number, we adopt the momentum representation by partial Fourier transform the expression (\ref{qintegral}) with respect to the first variable. With this in mind, we define the $s$-ordered quantization map of the function $g(c,\mu)$ as
\begin{equation}\label{LQCquantization}
\hat{g}_{s}\psi(c)=\mathcal{Q}^{LQC}_{s}(g)\psi(c):=\int_{\widehat{b\mathbb{R}}\times\widehat{b\mathbb{R}}}d\mu d\nu\,\tilde{g}\left( \frac{\mu-\nu}{1-s},\frac{\mu+\nu}{2}\right) h_{\frac{\mu-s\nu}{1-s}}(c)\tilde{\psi}_{\nu}, 
\end{equation}
where $\psi\in\Cyl(b\mathbb{R})$ and $s\in[-1,1]$. We can observe that for the case $s=0$, the formula (\ref{LQCquantization}) reduces to the Weyl-symmetric ordering quantization map determined in \cite{Sahlmann}. As an example, let us take the functions $g_{1}(c,\lambda)=h_{\lambda}(c)$ and $g_{2}(c,\lambda)=\lambda$, then by employing the $s$-ordered Weyl quantization prescription defined in (\ref{LQCquantization}) we obtain that their corresponding operators are given by
\begin{equation}
\mathcal{Q}^{LQC}_{s}(g_{1})\psi(c)=\hat{h}_{\lambda}\psi(c)
\end{equation}
and
\begin{equation}
\mathcal{Q}^{LQC}_{s}(g_{2})\psi(c)=\hat{p}\psi(c),
\end{equation}
for $\psi\in\Cyl(b\mathbb{R})$ and every value $s\in[-1,1]$. Further, if we now consider the function  $g_{3}(c,\lambda)=\lambda h_{\lambda}(c)$,  using the integration properties (\ref{haardRb}) and (\ref{haarRb}), we obtain the following results: for $s=0$, the map $\mathcal{Q}^{LQC}_{0}(g_{3})\psi(c)=\frac{1}{2}(\hat{h}_{\lambda}\hat{p}+\hat{p}\hat{h}_{\lambda})\psi(c)$, corresponds to the Weyl-symmetric ordering. For the value $s=1$, we obtain the standard ordering, $\mathcal{Q}^{LQC}_{1}(g_{3})\psi(c)=\hat{h}_{\lambda}\hat{p}\psi(c)$, and finally, when $s=-1$ the anti-standard ordering prescription holds $\mathcal{Q}^{LQC}_{-1}(g_{3})\psi(c)=\hat{p}\hat{h}_{\lambda}\psi(c)$.

\section{Conclusions}
\label{sec:conclu}
 
Within the Loop Quantum Cosmology framework, in this paper, we have generalized the Wigner-Weyl formalism by introducing an entire family of $s$-parametrized 
quasi-probability distributions defined in phase space, and their related $s$-ordered Weyl quantization maps.  Our construction  defines these quasi-distributions for states with values on the Bohr compactification of the real line, and  
closely follows the one proposed by  Cahill and Glauber in reference~\cite{Cahill}, where the parameter $s\in[-1,1]$ varies continuously,
thus labeling the normal and antinormal orderings in the extreme values of this interval while recovering the Weyl symmetric ordering for the value $s=0$. The main difference here, however, is related to the fact that since the quantum configuration space in the LQC representation  lies on the Hilbert space of square integrable functions on the Bohr compactification, a position operator simply does not exist.   In consequence, our definition of the  $s$-parametrized family of quasi-probability distributions for a state in the set of cylindrical functions $\Cyl(b\mathbb{R})$, must be carefully chosen in the momentum representation.  Indeed, in our prescription, we recover the Wigner distribution for LQC whenever we consider to set the value of the parameter $s$ to zero.  We also note that  the projections of the proposed $s$-parametrized quasi-distributions result in the well-known marginal probability densities, independently of the value of the parameter, thus implying their invariance under any ordering prescription.
Besides,  by evaluating  the $s$-ordered quasi-probability distribution for LQC for a single character $h_{\mu'}$, we have demonstrated that it determines a positive function
for an arbitrary ordering, as opposed to the  Schr\"odinger representation, where the only positive quasi-probability distributions are provided by Gaussian functions.  
Moreover, in order to define the $s$-ordered Weyl quantization map in LQC, we proceed similarly as the case introduced in~\cite{Cahill} by appropriately considering a distribution on the space $\Cyl(b\mathbb{R}\times\widehat{b\mathbb{R}})$.  In this sense, the proposed generalization of the Weyl quantization map for LQC not only reduces to its corresponding Weyl-symmetric ordering map counterpart for the value $s=0$ but also one may easily show that it contains the standard and anti-standard orderings for the appropriate values of the parameter $s$.

To conclude, we claim that both the $s$-parametrized 
quasi-probability distributions and the $s$-ordered Weyl quantization map described here
will be of a primordial relevance in order to study important issues within the LQC program, such as  correlations functions, squeezed states, and also the correspondence between the classical and quantum coherence theory, to mention some.  
We also claim that our developments will be helpful  to analyze the convergence of  operators and to the characterization of interference effects associated with quantum states, as examined recurrently in the analogous prescription 
inherent to  the quantum optics and the quantum information frameworks.  This will be reported elsewhere.

\section*{Acknowledgments}
The authors would like to acknowledge financial support from CONACYT-Mexico
under the project CB-2017-283838.

\section*{References}

\bibliographystyle{unsrt}

\end{document}